\documentclass[aps,reprint,floatfix,hidelinks,longbibliography,superscriptaddress]{revtex4-2}

\usepackage{graphicx}
\usepackage{dcolumn}
\usepackage{bm}
\usepackage[utf8]{inputenc}
\usepackage[T1]{fontenc}
\usepackage{mathptmx}
\usepackage{hyperref}
\usepackage[capitalize]{cleveref}  

\begin{document}


\title{Semiconductor membranes for electrostatic exciton trapping in optically addressable quantum transport devices}

\author{Thomas Descamps}
\affiliation{JARA-FIT Institute Quantum Information, Forschungszentrum Jülich GmbH and RWTH Aachen University, 52074 Aachen, Germany}
\author{Feng Liu}
\altaffiliation[Now at: ]{College of Information Science and Electronic Engineering, Zhejiang University, Hangzhou 310027, China}
\affiliation{JARA-FIT Institute Quantum Information, Forschungszentrum Jülich GmbH and RWTH Aachen University, 52074 Aachen, Germany}
\author{Sebastian Kindel}
\affiliation{JARA-FIT Institute Quantum Information, Forschungszentrum Jülich GmbH and RWTH Aachen University, 52074 Aachen, Germany}
\author{René Otten}
\affiliation{JARA-FIT Institute Quantum Information, Forschungszentrum Jülich GmbH and RWTH Aachen University, 52074 Aachen, Germany}
\author{Tobias Hangleiter}
\affiliation{JARA-FIT Institute Quantum Information, Forschungszentrum Jülich GmbH and RWTH Aachen University, 52074 Aachen, Germany}
\author{Chao Zhao}
\altaffiliation[Now at: ]{Key Laboratory of Semiconductor Material Science, Beijing Key Laboratory of Low Dimensional Semiconductor Materials and Devices, Institute of Semiconductors, Chinese Academy of Sciences, Beijing 100083, China}
\affiliation{Peter Grünberg Institute, Forschungszentrum Jülich GmbH, 52425 Jülich, Germany}
\author{Mihail Ion Lepsa}
\affiliation{Peter Grünberg Institute, Forschungszentrum Jülich GmbH, 52425 Jülich, Germany}
\author{Julian Ritzmann}
\affiliation{Lehrstuhl für Angewandte Festkörperphysik, Ruhr-Universität Bochum, D-44780 Bochum, Germany}
\author{Arne Ludwig}
\affiliation{Lehrstuhl für Angewandte Festkörperphysik, Ruhr-Universität Bochum, D-44780 Bochum, Germany}
\author{Andreas D. Wieck}
\affiliation{Lehrstuhl für Angewandte Festkörperphysik, Ruhr-Universität Bochum, D-44780 Bochum, Germany}
\author{Beata E. Kardyna\l}
\affiliation{Peter Grünberg Institute, Forschungszentrum Jülich GmbH, 52425 Jülich, Germany}
\affiliation{Department of Physics, RWTH Aachen University, 52074 Aachen, Germany}
\author{Hendrik Bluhm}
\affiliation{JARA-FIT Institute Quantum Information, Forschungszentrum Jülich GmbH and RWTH Aachen University, 52074 Aachen, Germany}
\email[]{bluhm@physik.rwth-aachen.de}

\date{\today}

\begin{abstract}
Combining the capabilities of gate defined quantum transport devices in GaAs-based heterostructures and of optically addressed self-assembled quantum dots could open broad perspectives for new devices and functionalities. For example, interfacing stationary solid-state qubits with photonic quantum states would open a new pathway towards the realization of a quantum network with extended quantum processing capacity in each node.
While gated devices allow very flexible confinement of electrons or holes, the confinement of excitons without some element of self-assembly is much harder. To address this limitation, we introduce a technique to realize exciton traps in quantum wells via local electric fields by thinning a heterostructure down to a 220 nm thick membrane.
We show that mobilities over $1 \times 10^{6}$ cm$^{2}$V$^{-1}$s$^{-1}$ can be retained and that quantum point contacts and Coulomb oscillations can be observed on this structure, which implies that the thinning does not compromise the heterostructure quality. Furthermore, the local lowering of the exciton energy via the quantum-confined Stark effect is confirmed, thus forming exciton traps. 
These results lay the technological foundations for devices like single photon sources, spin photon interfaces and eventually quantum network nodes in GaAs quantum wells, realized entirely with a top-down fabrication process.

\end{abstract}


\maketitle 
A tremendous amount of insight has been gained from a broad range of quantum transport experiments, ranging from the first observations of quantum point contacts and Coulomb blockade over quantum Hall physics all the way to the realization of spin qubits aspiring to become a highly scalable platform for quantum computing. For all of these topics, gateable two dimensional electron gases (2DEGs) in semiconductor heterostructures have been a workhorse. 
A similar wealth of results emerged from semiconductor (quantum) optics experiments using self-assembled quantum dots that accommodate confined exciton states. 
Yet, these domains have developed largely separately, partly due to the difficulty to electrostatically confine excitons. Notable exceptions include the detection of a the charge state of an ensemble of quantum dots via transport of a 2DEG or a doped back gate \cite{Kurzmann2016,Kurzmann2019} and the confinement of a small number of indirect excitons using gates \cite{Schinner2013}. One may thus expect very rich new possibilities from devices that combine both gateable carrier gases allowing the formation of quantum dots and controllably confined exciton states. 

A specific application of such a device, which is also the main motivation for this work, would be an interface between a matter qubit and a photonic qubit, which is an essential requirement to build a quantum communication network or a distributed quantum computer \cite{Kimble2008}. Over the last decade, major steps towards such networks have been made with optically-active solid-state qubits, in particular NV-centers in diamond \cite{Togan2010, Bernien2013, Pfaff2014} and self-assembled quantum dots (SAQDs) \cite{Gao2013, Schaibley2013, DeGreve2012}. While the photonic quantum resource can be exploited in remarkable ways \cite{Lodahl2018, Pezzagna2021}, using the spin resource of these qubits (electron/hole spin for SAQDs or Vacancy electron/Nitrogen-14 nuclear spin for NV centers in diamond) for quantum computing remains challenging \cite{Warburton2013, Gao2015, Pezzagna2021}.

For quantum computing nodes with at least tens of locally connected qubits, gate defined quantum dots (GDQD) formed in a semiconductor heterostructure are a potentially more compelling platform. While silicon is a natural choice for isolated quantum processors, devices in a GaAs/Al$_{x}$Ga$_{1-x}$As heterostructure have pioneered the field \cite{Hanson2007, Eizerman2004, Koppens2006,Foletti2009, Nowack2011, Shulman2012} and offer advantages for optical interfacing due to the direct band gap.
Yet the absence of hole confinement in a conventional electron-GDQD prevents the confinement of excitons and hence the coherent coupling to light. 
Earlier works have already demonstrated detection \cite{Fujita2013} and capture \cite{Oiwa} of a single photoelectron by a GDQD. However, the photohole could not be confined in these previous studies, leading to a loss of phase information. To remedy this limitation, one approach is to engineer the $g$-factor of electrons and holes to break their entanglement \cite{Kosaka2007, Kuwahara2010, Kosaka2011, Kuroyama2019}. Another idea pursued here is to add an optically active quantum dot (OAQD) able to confine excitons and acting as spin-photon interface in close vicinity of the GQDQ \cite{Joecker2018}. 

One candidate for an OAQD would be a SAQD grown above or below the two-dimensional electron gas \cite{Engel2006a}. These dots can be directly embedded into the III/V heterostructure and they were proven to be efficient single photon sources \cite{Warburton2013} or spin/photon interfaces \cite{Gao2012}. Tunnel coupling between an InAs SAQD and a reservoir has also been demonstrated \cite{Kurzmann2019}. However, all these major steps were achieved on SAQDs randomly distributed and having varying optical features due to the Stranski-Krastanov growth process. In other words, the dots showing the best features have to be manually found which limits the scalability of this approach. Besides, the strain introduced in the lattice by the SAQDs degrades the electron mobility of the 2DEG underneath. While values not much below $10^{6}$ cm$^{2}$V$^{-1}$s$^{-1}$ are desirable to reliably form GDQDs, the mobility of a 2DEG with embedded SAQDs typically remains one or two orders of magnitude lower \cite{Kurzmann2015}.

\begin{figure}[ht]
\centering
\includegraphics[width=\linewidth]{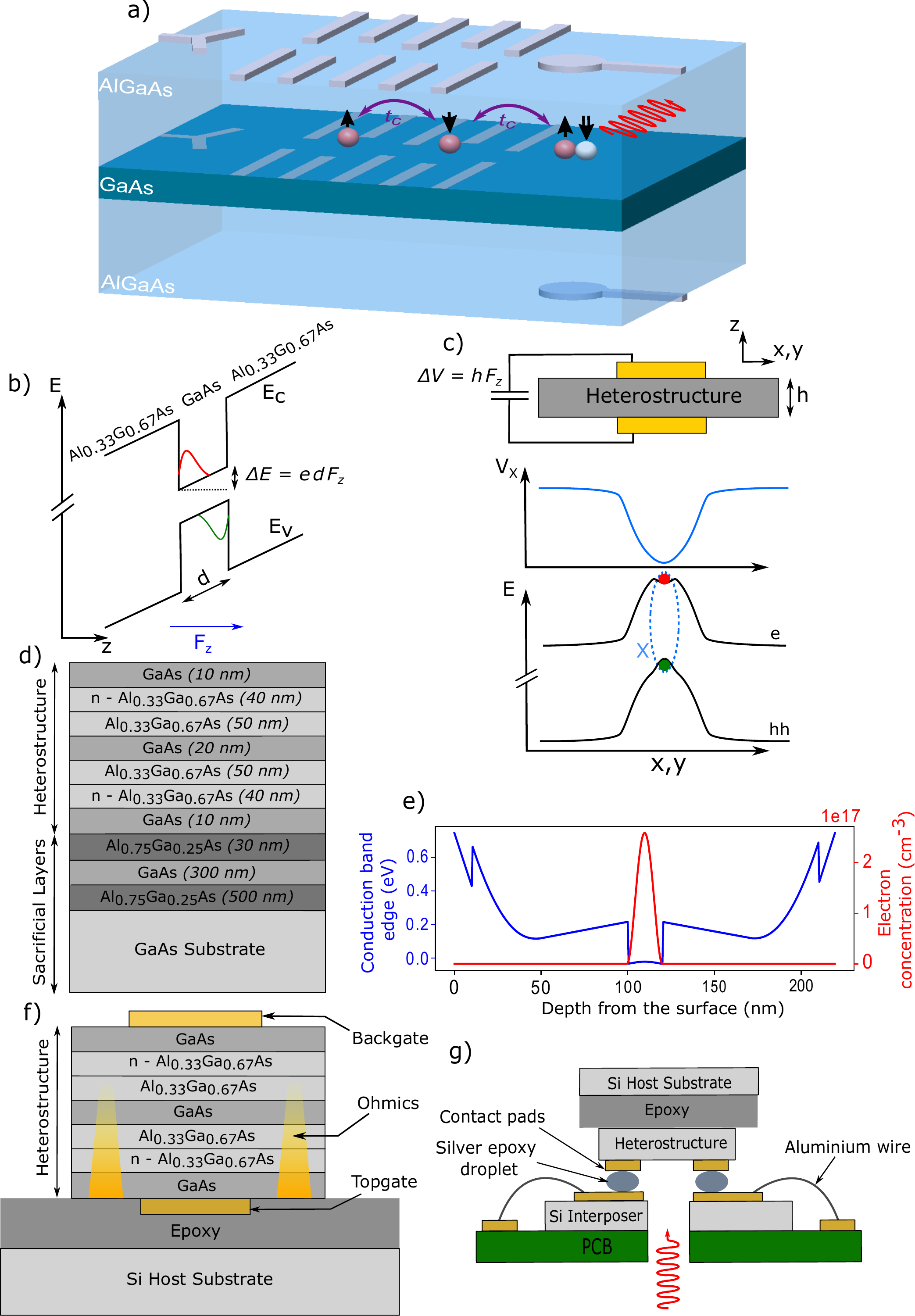}
\caption{\label{fig:sampleSkecth} a) Sketch of an optical interface for GDQDs (left side) based on an electrostatically-defined exciton trap (right side). The transfer of information is mediated by tunnel coupling $t_{c}$ b) Quantum-confined Stark effect on a Al$_{0.33}$Ga$_{0.67}$As/GaAs/Al$_{0.33}$Ga$_{0.67}$As heterostructure resulting from an electric field $F_{z}$ applied along the growth direction $z$. $E_{C}$ and $E_{V}$ are the conduction and valence band edges, respectively. The electron and heavy hole ground state wave functions inside the quantum well are depicted. c) In-plane spatial variation of electron and heavy hole energy bound by Coulomb interaction. Note that the variation of the mean of both quantities can be tuned independently of the difference induced by the electric field. The spatial variation of the effective potential $V_{X}$ of the resulting exciton $X$ is also sketched. d) Full layer stack composed of the sacrificial layers and the thin heterostructure investigated. e) Self-consistent simulation of the conduction band edge. f) Device layout for Hall measurements where the thin heterostructure is bonded to a silicon host with epoxy. Note that the original top surface is now in contact with the epoxy. g) Scheme to connect the contact pads on the device to a PCB.}
\end{figure}

Here, we introduce a top-down approach where the OAQD is fabricated deterministically and can be directly embedded in close vicinity to the GDQD to ensure tunnel coupling (\cref{fig:sampleSkecth}(a)). An additional expected advantage is the built-in electrical tunability and low spread of the operating wavelength.  
The idea to trap excitons is to use the quantum-confined Stark effect in a quantum well \cite{Dietl2019, Kowalik-Seidl2012, Schinner2011} in the following way (\cref{fig:sampleSkecth}(b)). A quantum well (QW) or two coupled quantum wells confine excitons along the growth direction. If an electric field is applied perpendicular to the quantum well, the exciton energy is lowered as the electron and hole can partially dissociate, thus creating an electric dipole interacting with the electric field. A complete dissociation is prevented by the confinement so that an overlap between the electron and hole wave function and thus optical addressability are retained. To achieve lateral confinement (\cref{fig:sampleSkecth}(c)), the electric field needs to be applied locally, which can be achieved with appropriately patterned gates. 
Such traps filled with only a few (indirect) excitons under sufficiently weak illumination have been demonstrated in a double quantum well gated by an unstructured, heavily doped back gate in the heterostructure and a patterned metal top gate \cite{Schinner2013}. 
One limitation of this specific realization is that a transverse electric field is necessarily associated with an in-plane field of similar magnitude due to the fixed potential at the back gate, which can laterally dissociate excitons. Besides, it is not possible to independently control the electric field determining the Stark shift and the electrostatic potential, which would be very advantageous for coupling confined exciton states to electrons in gate-defined quantum dots.


Here, we address these limitations by demonstrating a fabrication process and devices on a 200 nm thick membrane accommodating a quantum well with gates on both sides of the structure that were patterned using high resolution electron beam lithography. Due to the small size of the gates, a sufficient confinement to resolve the orbital level splitting between exciton states is expected. Electric potential and field can be controlled independently via the common and difference mode, respectively. Based on this device design, we demonstrate a local lowering of the exciton energy by up to 15 meV with 150 nm large gates (section \ref{sec:trap}).

A key ingredient of such structures is the removal for the GaAs substrate to pattern a metal gate on the bottom side of the heterostructure. This removal leaves an etched surface, which might degrade the quality of the heterostructure and cause detrimental charge noise. With our fabrication process (section \ref{sec:Design}), we show that mobilities similar to those required for single-electron quantum dots can be achieved (section \ref{sec:2DEG}) and demonstrate a quantum point contact and a quantum dot with high-quality characteristics in such a structure (section \ref{sec:pinchoff}). Our process thus overcomes key hurdles on the way to electrostatic exciton traps for devices like highly tunable single photon sources and spin-photon interfaces.


 


Similar device concepts have been used previously both for transport and optical experiments, using different approaches for their realization. 
One possibility is to grow a sacrificial layer between the substrate and the heterostructure, which is locally removed to create a suspended membrane. Transport studies have been carried out for this type of membrane hosting an electron gas \cite{Blick2000, Pogosov2012, Rossler2010}. Alternatively, the substrate is completely removed to fully expose the bottom side of the heterostructure. The second approach is more suited for our final purpose as it allows patterning of the two sides of the heterostructure with standard techniques. Independent contacts to two coupled quantum wells grown on a doped heterostructure have been demonstrated by having one contact on each side on the heterostructure \cite{Weckwerth1996}. The electron transport in the two quantum wells being coupled, resonant tunneling phenomena were studied with this method. Rather high mobilities of about $3.3 \times 10^{5}$ cm$^{2}$V$^{-1}$s$^{-1}$ (for an electron density between 1 and $4 \times 10^{11}$ cm$^{-2}$) in a doped \cite{Weckwerth1996} and nearly $10^{7}$ cm$^{2}$V$^{-1}$s$^{-1}$ (for an electron density of $5.8 \times 10^{11}$ cm$^{-2}$) in an undoped heterostructure were demonstrated \cite{DasGupta2012}. 
However, the thickness of the heterostructure and its gate stack was a few microns, which is much too large to achieve sufficiently tight exciton confinement and to realize single-electron quantum dots.
In the nanophotonics community, suspended membranes hosting self assembled quantum dots have been fabricated to create photonic cavities enhancing the light-matter interaction \cite{Laucht2012} or to couple the emitted light to photonic waveguides \cite{Javadi2018}. Complete removal of the substrate was also done to improve the collection efficiency by incorporating a metallic mirror underneath the heterostructure \cite{Wang2019a} or to tune the exciton emission with a piezoelectric substrate \cite{Ding2010}. Apart from the more conventional approach to confine excitons, these devices did not involve double sided high resolution patterning.


\section{\label{sec:Design}Design and fabrication of the devices}

Our overall strategy to fabricate a device on a thin semiconductor heterostructure follows the guidelines described by \citet{Weckwerth1996}. We optimized or adapted some steps to match our requirements. It starts with the patterning of the as-grown surface. The sample is then flipped and glued to a host Si substrate. The original GaAs substrate is completely etched away until the bottom cap and the resulting etched surface can finally be patterned.

The GaAs/Al$_{0.33}$Ga$_{0.67}$As modulation-doped heterostructure was grown on a 350 $\mu$m thick GaAs (100) substrate as illustrated on \cref{fig:sampleSkecth}(d). The first layers until the bottom GaAs cap consist of GaAs smoothing layers and Al$_{0.75}$Ga$_{0.25}$As sacrificial layers that will be etched away during the fabrication process. The remaining upper layers form the 220 nm thin heterostructure that will be investigated throughout this paper. The 20 nm undoped GaAs quantum well (QW) is separated from the silicon dopants ($6.5 \times 10^{17}$ cm$^{-3}$) in the Al$_{0.33}$Ga$_{0.67}$As barrier layers by a 50 nm spacer of intrinsic Al$_{0.33}$Ga$_{0.67}$As. \Cref{fig:sampleSkecth}(e) shows the conduction band profile simulated with a self-consistent Poisson-Schrödinger solver. The pinning level for the as-grown and the etched surface was assumed near mid-gap. The density and the position of the dopants was chosen to compensate for the surface states while still obtaining a high fraction of ionized dopants. 
Assuming that the Fermi level is pinned 120 meV below the conduction band edge in the doped region due to a high density of DX centers \cite{Mooney1990}, the expected carrier density is on the order $n_{s}=2.8 \times 10^{11}$ cm$^{-2}$ and only the lowest sub-band is populated.

The final sample architecture of an exemplary Hall bar with a top and bottom Schottky gate is shown in \cref{fig:sampleSkecth}(f). Conventional photolithography is used to pattern a Hall bar on the topside. A mesa (not shown) is etched down to at least 210 nm, thus removing the complete heterostructure, with a 100:3:3 H$_{2}$O/H$_{2}$O$_{2}$/H$_{3}$PO$_{4}$ solution and AuGeNi ohmics are metallized and annealed at 480$^{\circ}$C for 60 s in a N$_{2}$/H$_{2}$ atmosphere. The top Schottky gate is then patterned on top of the mesa and a semi-transparent metal stack (2 nm Ti/7 nm Au) was deposited. The topside of the resulting Hall bar is then glued with a thin layer of low-curing temperature and low-stress epoxy (Epotech 353ND) to a Si host substrate before removal of the GaAs substrate. This epoxy shows the appropriate viscosity to be dispensed as a small drop on the sample. The host substrate is manually pressed on top with a low force and slid back and forth to spread the epoxy over the surface of the sample. The resulting thickness of the squeezed epoxy is not precisely controlled but is typically around a few microns. In principle, any type of substrate can be used as a host as long as the epoxy has a good adhesion on it, but we found silicon convenient as it is not damaged by the chemicals used in the process. An etching solution (H$_{2}$O$_{2}$+citric acid) selectively etches the whole GaAs substrate until the 500 nm Al$_{0.75}$Ga$_{0.25}$As etch stop which is not attacked \cite{Kim1998}. It was prepared from citric acid monohydrate dissolved 1:1 by weight into deionized water and H$_{2}$O$_{2}$ was added so that the overall mixing ratio citric acid/H$_{2}$O$_{2}$ was 1:4 in volume. This ratio turned out to be a good compromise between fast etch rate ($\approx$ 7 nm/s) and good selectivity. The etch stop is removed by a 1\% diluted hydrofluoric (HF) solution ($\approx$ 8 nm/s). Fine particles typically remain on the surface, and they are speculated to be a hydroxide of aluminium \cite{Khankhoje2010} formed during the etch stop removal. They can be dissolved by a 30 s dip in a hydroxide potassium solution (25 g/100 ml dionized water). Our experience showed that a full chemical etching of the substrate instead of a combination mechanical polishing/chemical etching \cite{Weckwerth1996, DasGupta2012} reduces the risk of cracks on the thin semiconductor heterostructure. Another cycle of (H$_{2}$O$_{2}$+citric acid) + HF etches the 300 nm GaAs buffer layer and the 30 nm Al$_{0.75}$Ga$_{0.25}$As second etch stop. These are not fundamentally necessary for the fabrication process, but a two-step substrate removal consisting of a first thick etch stop followed by a thinner one is expected to provide a smoother surface at the bottom GaAs cap. Once the latter is accessible, the backgate can be patterned in the same way as the top gate. The alignment accuracy achieved with optical lithography between the top and bottom features is a few microns as the 220 nm heterostructure is sufficiently transparent. Finally, holes are etched through to access the contact pads of the buried ohmics and topgate. Following all these steps, we could reach yields around 80\% for Hall Bars. The most common failure mode is physical damage of the sample during processing.
For the other devices discussed in sections \ref{sec:pinchoff} and \ref{sec:trap}, the nanometer-scale gates were written with electron beam lithography. Two thicknesses (2 nm Ti/ 7 nm Au or 5 nm Ti/ 20 nm Au) for these gates were considered depending on the transparency requirement. The alignment of the top and bottom gates used for the exciton trap was simplified by the transparency of the thinned heterostructure: the markers written on the top side could be imaged on the bottom side after flipping. On the SEM micrograph of the exciton trap in \cref{fig:starkshift}(a), the top gates located under the flipped heterostructure (blurry edges) and the bottom gates located on top (sharp edges) were imaged simultaneously. While spatially separated far from the trap region, the top and bottom gates overlap near the exciton trap (inset of \cref{fig:starkshift}(a)) with an alignment accuracy of $\pm 35$ nm. To obtain this estimate, two concentric metal disks were placed on both surfaces (2.5 $\mu$m radius/ 9 nm thick on the bottom surface and 1 $\mu$m radius/ 25 nm thick on the top surface). The measured center-to-center distance between the two disks reflects the misalignment.

Conventional wire or ball bonding techniques were not successful to connect the ohmics and the gates of our samples to a chip carrier due to the softness of the membrane and the epoxy. The bonding technique that we opted for is sketched on \cref{fig:sampleSkecth}(g). Fine droplets of silver epoxy are placed on the contact pads of a silicon interposer either manually or with a stencil. The 220 nm heterostructure is then flipped and pressed onto a Si interposer with a flip-chip bonder. The silver epoxy is baked for 5 minutes on the bonder while the interposer and the membrane are aligned. A longer curing (20 min) then takes place in an oven to strengthen the bond. The interposer can finally be connected to a PCB with a wire bonder. Compared to the manual gluing of a wire with silver epoxy between the contact pads of the sample and chip carrier, our approach is more reliable and can easily connect a large number of pads. Finally, to illuminate the surface of the sample, an opening is made through the PCB and the silicon interposer. The PCB is simply drilled and a 3x3 mm hole is cut on the interposer with a laser marker (Coherent CombiLine Advanced). Typical writing parameters were: 33 A driving current of the laser head, 15 kHz pulse repetition rate and 50 mm/s scanning speed. The surface of the interposer was protected beforehand with a thick layer of photoresist. The accuracy on the position of the hole was on the order of a few hundred of microns, which is good enough for our purpose. We found the laser cutting much easier and quicker than wet or dry etching of silicon. Compared to a glass interposer, an open window in silicon avoids transmission and reflection losses and ensures that the thermal expansion of the interposer is matched to that of the host substrate. 

\section{\label{sec:2DEG}Electron gas properties on thin semiconductor heterostructure}

The Hall bar described above was characterized electrically and optically to extract the carrier density $n_{s}$ and the mobility $\mu$ of the 2DEG. Only the lowest sub-band of the 2DEG is supposed to be populated and the device should be gateable without hysteresis.
\begin{figure}[ht]
\centering
\includegraphics[width=\linewidth]{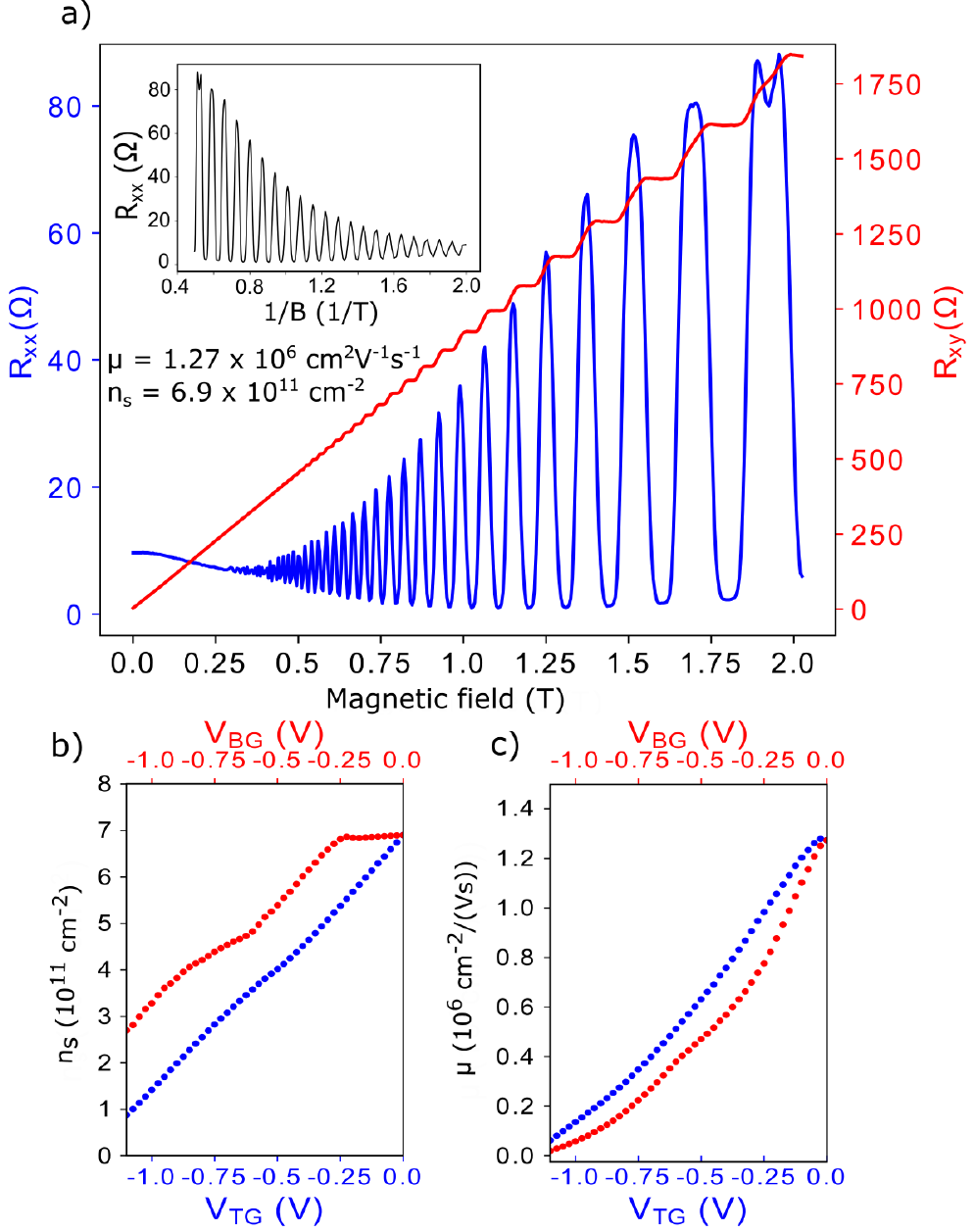}
\caption{\label{fig:Hall} a) Hall resistance $R_{xy}$ and Shubnikov de-Haas oscillations for $V_{TG}$ = $V_{BG}$ = 0 V. The inset shows the clear $1/B$ period of the oscillations. The carrier density $n_{s}$ (b) and the mobility $\mu$ (c) can be tuned with the topgate and backgate voltage.}
\end{figure}
The Hall bar was cooled to 35 mK in a dilution refigerator equipped with an optical window, which corresponds to the intended operating conditions for the exciton traps and qubits coupled to it. An objective lens was mounted on the sample stage for confocal microscopy. We performed Hall measurements with a standard 4-terminal method with a driving current of 30 $\mu$A without illumination. \Cref{fig:Hall}(a) shows the longitudinal and transverse voltage when the topgate voltage ($V_{TG}$) and backgate voltage ($V_{BG}$) are set to zero. Clear Shubnikov-de Haas (SdH) oscillations can be seen and the ($1/B$) spectrum (inset \cref{fig:Hall}a) shows a single period. Note that the splitting of the peaks starting to occur around 2 T corresponds to the lifting of the spin degeneracy. Hence, the conduction channel consists of only the lowest subband. From the minima of the oscillations, the carrier density of the 2DEG is $n_{s} = 6.9 \times 10^{11}$ cm$^{-2}$ and the mobility reaches $\mu = 1.27\times 10^{6}$ cm$^{2}$V$^{-1}$s$^{-1}$. An identical carrier density was found from the slope of the Hall voltage as a function of magnetic field. As a reference, the same heterostructure as \cref{fig:sampleSkecth}(a) was grown in the same MBE chamber without the sacrificial layers to avoid an undesired electron channel at the interface between the 30 nm Al$_{0.75}$Ga$_{0.25}$As etch stop and the 300 nm GaAs buffer layer. A Hall bar with identical geometry, was made on this wafer, however without etching away the GaAs substrate. At 4.2 K, the carrier density was $n_{s} = 7.1 \times 10^{11}$ cm$^{-2}$ and the mobility $\mu = 1.51 \times 10^{6}$ cm$^{2}$V$^{-1}$s$^{-1}$. Keeping in mind the different measurement temperatures and the growth fluctuations, we can conclude that the heterostructure was not significantly damaged after the substrate removal and further processing. The mobility of the thinned heterostructure remains above $10^{6}$ cm$^{2}$V$^{-1}$s$^{-1}$, which is commonly considered sufficient to form single-electron quantum dots. 
In both samples, the carrier concentrations measured are larger than the simulated value. This discrepancy 
may be due to an insufficient concentration of DX centers in the doped barriers \cite{Mooney1990a}, which leads to pinning of the Fermi level just a few meV below the conduction band due to shallow, unstable traps. 
The response of the carrier density to the topgate and backgate (\cref{fig:Hall}(b)) is consistent with this hypothesis. While the topgate can directly deplete the gas, the backgate starts to have an effect only after a threshold voltage of $-0.25$ V. This behavior may be due to the filling of shallow traps or excess surface states on the etched surface, either of which would screen the backgate voltage. The high carrier density and the fact that the mobility in \cref{fig:Hall}(c) declines even at fixed density suggests that shallow traps are the more likely cause. 
We thus expect that fine tuning of the growth parameters can improve the gateability and stability of the device, which is backed up by results from quantum point contacts on a different heterostructure discussed below.

\begin{figure}[ht]
\centering
\includegraphics[width=\linewidth]{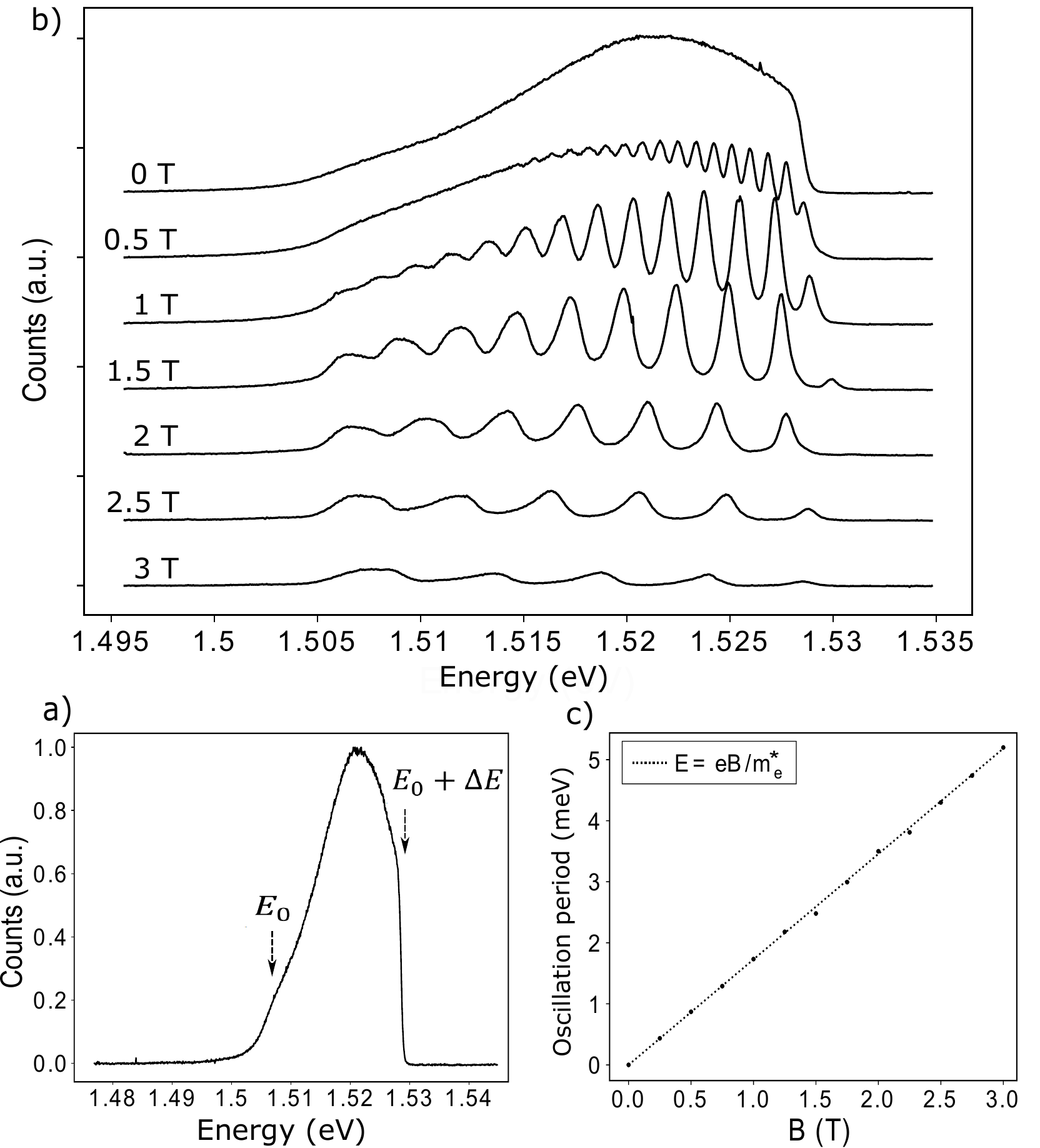}
\caption{\label{fig:oscillations+fit} a) PL lineshape of the double side doped heterostructure. $E_{0}$ corresponds to the bottom of the quantum well energy band and the Fermi edge can be seen at $E_{0} + \Delta E$. b) PL spectra at different magnetic fields. c) Period of the oscillations as a function of the perpendicular magnetic field. The dashed line corresponds to the theoretical cyclotron energy.}
\end{figure}

The single subband nature of the 2DEG can also be confirmed by photoluminescence measurements under magnetic fields. The sample was illuminated by a 795 nm continuous laser. The excitation energy is below the Al$_{0.33}$Ga$_{0.67}$As energy barrier to avoid the creation of carriers in the barrier. The spot size on the sample was approximately 3 $\mu$m with an average power of 100 nW. The photoluminescence (PL) was dispersed by a 1 m grating spectrometer and detected by a CCD thermo-electrically cooled to $-80 ^{\circ}$ C. 
\Cref{fig:oscillations+fit}(a) shows the PL spectrum at zero magnetic field. The asymmetric and broad line shape exhibits the characteristic features of a QW hosting an electron gas \cite{Kamburov2017, Skolnick1987}: the bottom of the quantum well energy band at $E_{0} = 1.507$ eV and the Fermi edge at higher energy $E_{0} + \Delta E = 1.527$ eV. The absence of a shoulder near the Fermi-edge shows that the second conduction subband is not occupied \cite{Pinczuk1984, Chen1991}. The energy $E_{0}$ is redshifted by approximately 15 meV compared to the energy measured at 8 K for a 20-nm quantum well on GaAs substrate \cite{Polland1985}. In addition to lower temperature, we attribute this difference by a change of strain on the quantum well after substrate removal \cite{Mei2007}. The overall width of the emission $\Delta E$ gives an estimate of the Fermi energy \cite{Pinczuk1984, Skolnick1987, Kamburov2017} $E_{F}= \Delta E / (1+m_{e}^{*}/m_{h}^{*}) = 23$ meV, where $m_{e}^{*}$ ($m_{h}^{*}$) is the effective electron (hole) mass. Assuming the 2D electron density of states of GaAs $g = 2.8 \times 10^{10}$ meV$^{-1}$cm$^{-2}$, the corresponding carrier density is $n=gE_{F}=6.48\times10^{11}$ cm$^{-2}$, which is closed to value found from the  transport measurement discussed above. Upon applying a fixed perpendicular magnetic field (\cref{fig:oscillations+fit}(b)), the PL spectra exhibit periodic oscillations on the high energy side corresponding to the different Landau levels being populated \cite{Skolnick1987}. The period of the oscillations as a function of the magnetic field is plotted on \cref{fig:oscillations+fit}(c) with the theoretical cyclotron energy $\omega_{e}=eB/\hbar m_{e}^\ast$ corresponding to the energy spacing between two consecutive Landau levels of the first subband. The energies found experimentally are very close to the expected values, which confirms that only the lowest sub-band is occupied.

\section{\label{sec:pinchoff}Quantum transport measurements}

\begin{figure}
\centering
\includegraphics[width=\linewidth]{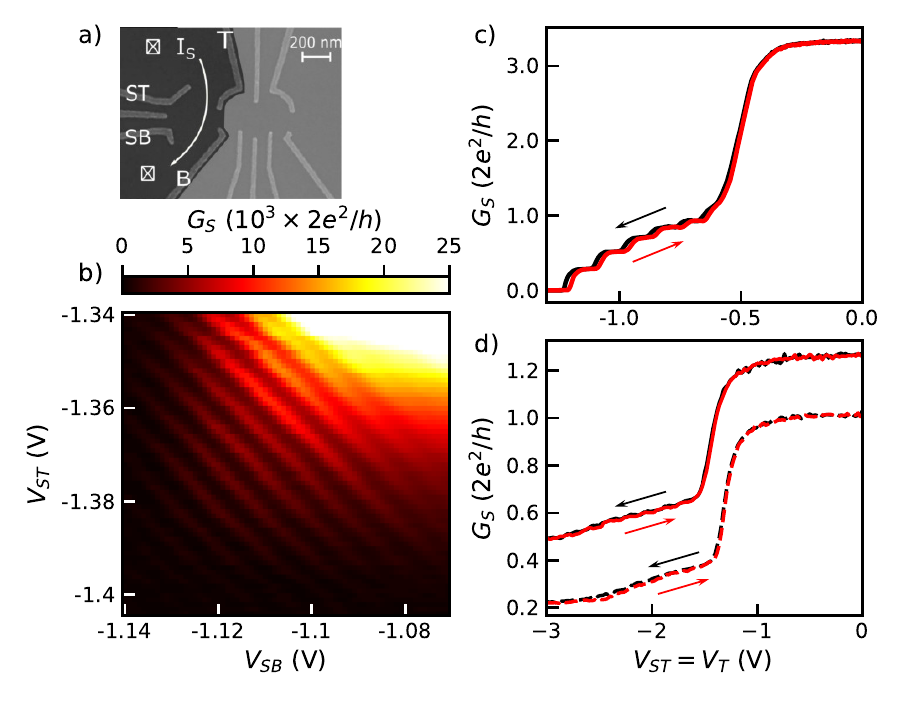}
\caption{\label{fig:pinch-off} a) Scanning electron micrograph of the device. Only the left part of the device is used to form a quantum dot, which would be used for charge sensing when operating the full device. b) Sensor dot conductance for fixed gate voltages $V_{T} = -1.5$ V and $V_{B} = -1.4$ V showing clean Coulomb oscillations. c) Conductance $G_{S}$ through the sensor dot when the two gates ST and T were swept symmetrically ($V_{SB} = V_{B} = 0$ V). d) Sweep as in c) but on a different wafer using gates on the unetched (solid lines, offset by $G_0/4$ for clarity) and etched (dashed lines) side of the heterostructure. No qualitative difference can be seen, indicating the etching does not harm the transport properties of the structure.}
\end{figure}

To complement the assessment of the heterostructure quality with a Hall bar and to verify the suitability for quantum transport experiments, we fabricated a GDQD on the thinned heterostructure membrane. Depletion Ti/Au (5 nm Ti, 20 nm Au) gates were patterned on the topside as shown in \cref{fig:pinch-off}(a). The device was cooled to 35 mK and only the left part (sensor dot) was investigated. The conductance was measured with a standard low-frequency lock-in technique with an AC source-drain voltage between 80 and 150 $\mu$V. \Cref{fig:pinch-off}(b) shows the conductance through the sensor dot at fixed voltages $V_{T}$ and $V_{B}$ as a function of $V_{SB}$ and $V_{ST}$, where the subscripts refer to the gate names shown in \cref{fig:pinch-off}(a). In this regime, the device consists of a single large dot as Coulomb oscillations can be identified clearly. When the T and ST gate voltages $V_{T}$ and $V_{ST}$ are swept symmetrically with all other gates grounded (\cref{fig:pinch-off}(c)), clear plateaus resulting from the quantization of the conductance can be resolved, which demonstrate that the two split gates behave as a quantum point contact. 
The position of the plateaus deviates from the integer values of $G_{0} = 2e^{2}/h$ due to measurement artefacts associated with the two-terminal measurement used.

To further investigate the influence of etching the heterostructure down to a thin membrane on the transport properties, the same gate pattern was fabricated once on the unetched side and once on the etched side of the heterostructure on a different wafer. The gate structures were offset in-plane from each other by a few microns. \Cref{fig:pinch-off}(d) shows the pinch-off behavior for the unetched (solid lines) and etched (dashed lines) sides for the lithographically same pair of gates $V_{ST}$ and $V_{T}$ measured at 4 K. The curves are offset by $G_0/4$ for clarity. While on this particular wafer, the channel could not be completely pinched off even with much more negative voltages, there is no discernible difference between the data for the front- and backsides of the membrane for all gate pairs of the devices, indicating that the etched surface does not deteriorate the electrical properties of the sample, for example due to increased interface defects. In particular, these measurements as well as reference measurements on other heterostructures confirm that signs of hysteresis seen in Figs.~\ref{fig:Hall}(b),~\ref{fig:pinch-off}(c) and other measurements on the same structure are related to the hetererostructure itself rather than a consequence of the etching process.

\section{\label{sec:trap}Exciton trap}

\begin{figure}[ht]
\centering
\includegraphics[width=\linewidth]{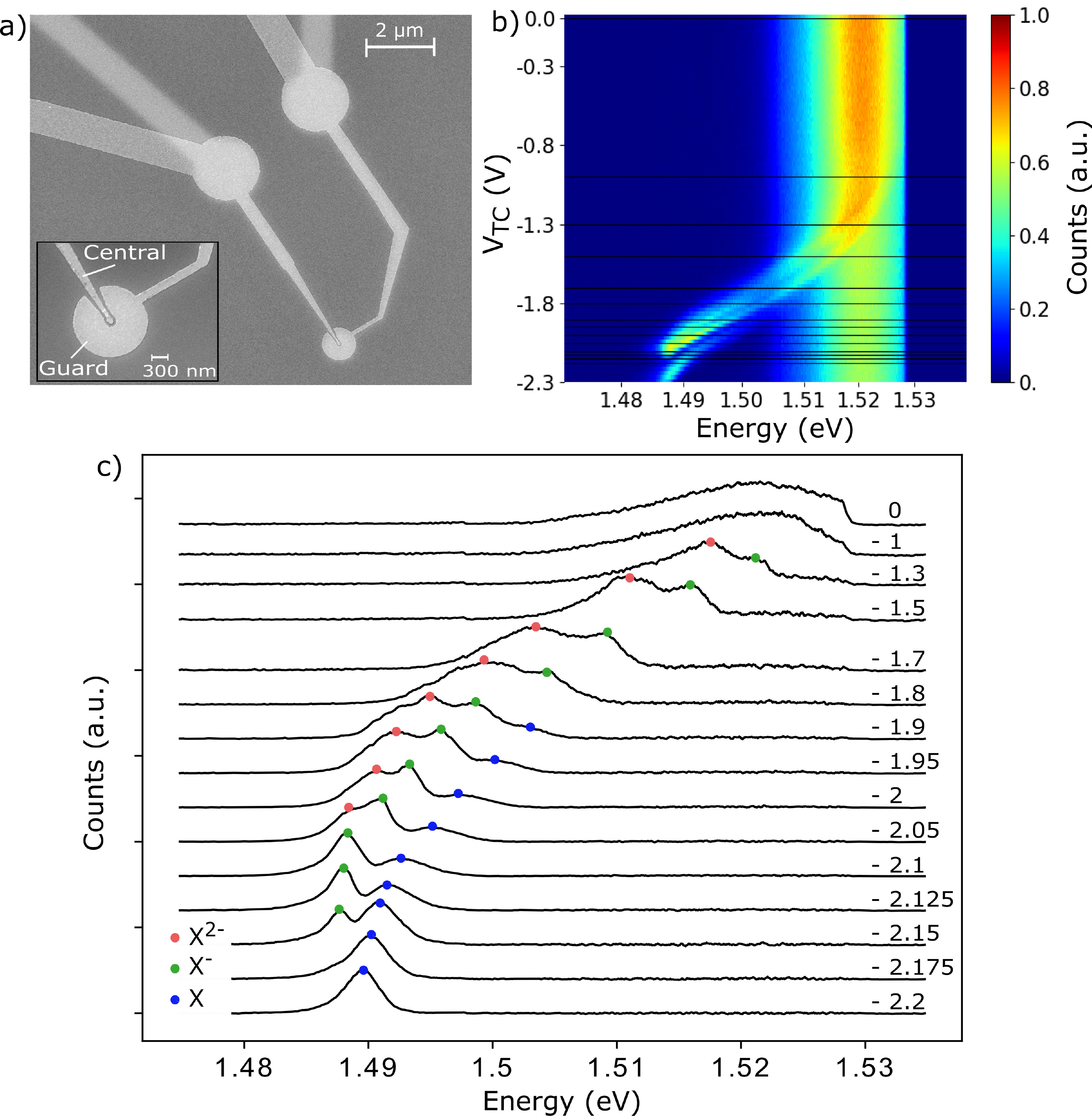}
\caption{\label{fig:starkshift} a) A scanning electron image of the central gates and the guard gates used to create the exciton trap. The same gates were also patterned on the top and bottom surface. b) PL measurement at different top central gate $V_{TC}$ voltage. All the other gates of the exciton trap were set to 0 V. c) Normalized PL spectrum extracted from b) (horizontal black lines) at various top central gate $V_{TC}$. Each spectrum has been shifted for clarity. The dots on the spectra identify the different exciton lines. The PL emission of the unbiased QW outside the metal gates has been subtracted to all the spectrums.}
\end{figure}

The second key ingredient to implement an electrostatic exciton trap is the spatially localized electric field, which requires patterning of the gates on a much shorter length scale than in a hall bar. To realize optical coupling to a gate-defined spin qubit, for example using the scheme described by \citet{Joecker2018}, this optically active quantum dot (OAQD)  will ultimately have to be embedded next to the GDQD. \Cref{fig:starkshift}(a) shows the geometry of an isolated trapping dot. On each side of the heterostructure, two semi-transparent metal gates are fabricated by e-beam lithography: one circular gate called central gate (diameter 150 nm) and one surrounding ring gate called the guard gate (diameter 1500 nm) separated by a 50 nm gap. The top central (guard) gate is vertically aligned with the bottom central (guard) gate. The central gates are used to confine excitons underneath them by the quantum-confined Stark effect. The trapping potential is expected to be approximately parabolic in the small region of interest due the distance of the gates from the 2DEG. The guard gates cause a more rapid reduction of the electric field outside the region between the central gates.

The sample was characterized optically at 35 mK with the same optical setup as described in section \ref{sec:2DEG}. It was excited at 790 nm with an average power of 300 nW on the surface. \Cref{fig:starkshift}(b) shows the PL spectra measured as a function of the voltage applied to the top central gate $V_{TC}$ with all other gates grounded. Until $V_{TC} = -1$ V, the gate voltage does not noticeably affect the QW emission energy centered around 1.525 eV. Despite the expected gradual depletion of the 2DEG, the electron density remains large enough in this range to screen the electric field so that the Stark effect is  suppressed.  
As the gate voltage increases from $-1$ V ($n_{s} \approx 1 \times 10^{11}$ cm$^{-2}$ according to \cref{fig:Hall}(b)), the electron carrier density keeps decreasing to the stage where a transition from a populated to an empty well occurs. As a result, a strong Stark shift is observed.
Note that the unbiased QW is always visible as well since the laser spot size is larger than the diameter of the top central gate.

To further examine the Stark shift effect, individual spectra from \cref{fig:starkshift}(b) are plotted in \cref{fig:starkshift}(c). Here, the PL of the unbiased quantum well outside the metal gate has been subtracted for clarity. At $-2.1 < V_{TC} < -1.90$ V, three exciton lines gradually shifting to lower energies by the quantum-confined Stark effect can be identified. 
Following earlier polarization and photoluminescence studies carried out by \citet{Finkelstein1995a} on a doped GaAs quantum well, we propose the following  peak identification. The carrier density being still large in that range, the heavy-hole exciton bound to two electrons X$^{2-}$ and the heavy-hole exciton bound to one electron also called negative trion X$^{-}$ dominate the spectrum. The heavy-hole neutral exciton X is only weakly observed. From $V_{TC} < -2.1$ V, the doubly charged exciton X$^{2-}$ disappears and only the trion X$^{-}$ and the neutral exciton X remain distinctly separated on the spectrum. The X$^{-}$ trion line originally dominates the neutral exciton X and at $V_{TC} = -2.15$ V, the intensity of the two peaks becomes comparable. From $V_{TC} < -2.175$ V, the trion peak vanishes and the neutral exciton peak eventually dominates the spectrum, showing that the 2DEG has been completely depleted. According to \cref{fig:Hall}, we would expect the electron gas to be depleted around $-1.5$ V. The depletion occurs at larger voltages in \cref{fig:starkshift} due to partial screening of the central gate voltage by the guard gate.
At $V_{TC} = -2.2$ V, the neutral exciton X line is red-shifted from the low energy tail of the unbiased QW by 15 meV. A spatially-localized exciton trap with a confinement potential of 15 meV is therefore formed under the central gate. Scanning-PL experiments on similar devices have also revealed the expected spatial variation of the peak structure, however the optical resolution is insufficient to properly resolve the trapping potential.
Similar behaviors are obtained when the bias is applied on the top central gate. 

\section{\label{sec:conclusion}Conclusion}
In summary, we describe a process that allows the fabrication of fine gate structures with ebeam lithography on both sides of a 220 nm thick membrane containing a GaAs/AlGaAs heterostructure. 
We show that removing the original GaAs substrate using wet etching and etch stops does not deteriorate the electrical and optical properties of the 2DEG hosted. Quantum point contacts and GDQDs with clean signatures of Coulomb blockade can be formed. Using an appropriate gate pattern on both sides, the exciton energy can be lowered locally by up to 15 meV.
These results lay the technological foundations for more advanced experiments and devices uniting aspects from the field of quantum transport and semiconductor (quantum) optics. One specific motivation is the realization of a spin-photon interface for qubits hosted in gate-defined quantum dots. On the transport side, this will require the realization of single-electron quantum dots and qubit operation in membrane devices. The next steps for optical experiments are the verification of single-photon source properties and level structure of the exciton traps to confirm that an optically active quantum dot can indeed be formed. Subsequently, more complex devices combining both aspects must be realized. 



\begin{acknowledgments}
This work was funded by the European Research Council (ERC) under the European Union's Horizon 2020 research and innovation program (Grant agreement No. 679342), by the Deutsche Forschungsgemeinschaft (DFG, German Research Foundation) under Germany's Excellence Strategy - Cluster of Excellence Matter and Light for Quantum Computing (ML4Q) EXC 2004/1 -- 390534769, and by the Deutsche Forschungsgemeinschaft (DFG, German Research Foundation) -- 328514164.

We acknowledge the support of the cleanroom Helmholtz Nano Facility (HNF) at Forschungszentrum Jülich, in particular Florian Lentz and Stefan Trellankamp for their support with the electron beam lithography. Besides, we thank the Institut für Werkstoffe der Elektrotechnik (IWE1) at RWTH Aachen University for giving us access to their flip-chip bonder and the Institute für Halbleitertechnik (IHT) at RWTH Aachen University for allowing us to use their laser marker.

C.Z., M.I.L., J.L., A.L., A.D.W. grew prototype wafers and in particular the wafer used for this project. T.D. fabricated the devices. T.D. and R.O. developed the bonding technique. T.D. and F.L. designed the cryostat. T.D. measured the samples. T.H. performed additional transport measurements. T.D. and S.K. and H.B. analyzed the data. H.B. and B.E.K provided guidance on various aspects of the experiments. T.D. and H.B. wrote the manuscript with input from other authors.
\end{acknowledgments}

\bibliography{References.bib}

\end{document}